\documentclass[apj,twocolappendix, numberedappendixn]{emulateapj}
\usepackage[colorlinks,citecolor=blue]{hyperref}
\usepackage{textcomp}
\usepackage{subfigure}
\usepackage{verbatim}
\usepackage{bm}% bold math
\usepackage{hyperref}
\usepackage{amsmath}
\usepackage{graphicx}
\usepackage{epstopdf}
\usepackage{amssymb}
\usepackage{extarrows}
\usepackage{color}
\usepackage{CJK}
\usepackage{cancel}
\usepackage{ulem}

\newcommand{\ba}{\begin{eqnarray}}
\newcommand{\ea}{\end{eqnarray}}
\newcommand{\be}{\begin{equation}}
\newcommand{\ee}{\end{equation}}
\newcommand{\gr}{\mathrm{GR}}

\newcommand{\m}{\mathrm{max}}
\newcommand{\mi}{\mathrm{min}}

\newcommand{\oct}{\mathrm{oct}}

\newcommand{\au}{\mathrm{AU}}

\newcommand{\IN}{\mathrm{in}}
\newcommand{\OUT}{\mathrm{out}}

\newcommand{\lk}{\mathrm{LK}}
\newcommand{\merger}{\mathrm{merger}}
\newcommand{\eff}{\mathrm{eff}}
\newcommand{\f}{\mathrm{f}}
\newcommand{\SL}{\mathrm{sl}}

\newcommand{\tot}{\mathrm{tot}}

\def\e1{e_1^2}

\begin{document}
\title{Binary Mergers near a Supermassive Black Hole: Relativistic Effects in Triples}
\author{Bin Liu$^{1}$, Dong Lai$^{1,2}$, Yi-Han Wang$^{3}$}
\affil{$^{1}$ Cornell Center for Astrophysics and Planetary Science, Cornell University, Ithaca, NY 14853, USA\\
$^{2}$ Tsung-Dao Lee Institute, Shanghai Jiao Tong University, Shanghai 200240, China\\
$^{3}$ Department of Physics and Astronomy, Stony Brook University, Stony Brook, NY 11794-3800, USA
}

\begin{abstract}
We study the general relativistic (GR) effects induced by a spinning
supermassive black hole on the orbital and spin evolution of a merging
black hole binary (BHB) in a hierarchical triple system.  A
sufficiently inclined outer orbit can excite Lidov-Kozai eccentricity
oscillations in the BHB and induce its merger.  These GR effects
generate extra precessions on the BHB orbits and spins, significantly
increasing the inclination window for mergers and producing a wide
range of spin orientations when the BHB enters LIGO band. This
``GR-enhanced" channel may play an important role in BHB mergers.
\end{abstract}

%  \rightharpoondown
%\maketitle

\section{Introduction}

The detections of gravitational waves from merging binary black holes (BHs)
\citep[e.g.,][]{Abbott 2018a,Abbott 2018b,Zackay 2019,Venumadhav 2019}
have motivated many recent studies on the dynamical formation of such
compact black-hole binaries (BHBs). Dynamical formation channels
include mergers arising from strong gravitational scattering in dense
clusters \citep[e.g.,][]{Zwart (2000),OLeary (2006),Miller (2009),Banerjee (2010),Downing (2010),Ziosi (2014),Samsing (2017),Samsing (2018a),Samsing (2018b),Rodriguez (2018),Gondan (2018)}
and more gentle ``tertiary-induced mergers'' -- the
latter can take place either in isolated triple/quadrupole systems
\citep[e.g.,][]{Antonini (2017),Silsbee (2017),Liu-ApJL,Liu-ApJ,Liu spin}
or in nuclear clusters dominated by a central supermassive BH (SMBH)
\citep[e.g.,][]{Antonini 2012,VanLandingham 2016,Petrovich 2017,Hoang 2018,Hamers 2018,Xianyu 2018,Fragione SMBH}.

In this paper we are interested in stellar-mass BHB mergers induced by a SMBH.
Such BHBs may exist in abundance in the nuclear cluster around the SMBH
due to various dynamical processes, such as scatterings and mass segregation
\citep[e.g.,][]{OLeary 2009,Leigh 2018}. Gravitational perturbation from
the SMBH induces Lidov-Kozai (LK) eccentricity oscillations
of the BHB, which leads to enhanced gravitational radiation and merger of the BHB.
Our paper examines several general relativistic (GR) effects that are overlooked
in previous studies, but significantly impact the efficiency and outcomes of LK-induced mergers.
We focus on isolated BHB-SMBH systems, and do not consider other processes related to
scatterings and relaxation with surrounding stars in the cluster \citep[e.g.,][]{VanLandingham 2016,Petrovich 2017,Hamers 2018},
which may also change the character of LK-induced mergers.

In the {\it Standard LK-Induced Merger} scenario,
a BHB with masses $m_1$, $m_2$, semimajor axis $a_\IN$ and eccentricity $e_\IN$,
moves around a tertiary ($m_3$) on a wider orbit with $a_\OUT$ and $e_\OUT$.
The angular momenta of the inner and outer binaries are
denoted by $\textbf{L}_\IN\equiv\mathrm{L}_\IN\hat{\textbf{L}}_\IN$ and
$\textbf{L}_\OUT\equiv\mathrm{L}_\OUT\hat{\textbf{L}}_\OUT$ (where $\hat{\textbf{L}}_\IN$ and $\hat{\textbf{L}}_\OUT$ are unit vectors).
If the mutual inclination between $\hat{\textbf{L}}_\IN$ and
$\hat{\textbf{L}}_\OUT$ (denoted as $I$) is sufficiently high, the inner
binary would experience LK eccentricity oscillations on the timescale
%%%%%%%%%%%%%%%%%%%%%%%%%%%%%%%%%%%%%%%%%%%%%%%%%%%%%%%%%%%%%%%%%%%%%%
\be
t_\lk=\frac{1}{\Omega_\lk}=\frac{1}{n_\IN}\frac{m_{12}}{m_3}\bigg(\frac{a_{\OUT,\eff}}{a_\IN}\bigg)^3,
\ee
%%%%%%%%%%%%%%%%%%%%%%%%%%%%%%%%%%%%%%%%%%%%%%%%%%%%%%%%%%%%%%%%%%%%%%
where $m_{12}\equiv m_1+m_2$, $n_\IN=(G m_{12}/a_\IN^3)^{1/2}$ is the mean motion of the inner binary,
and $a_{\OUT,\eff}\equiv a_\OUT\sqrt{1-e^2_\OUT}$ is the effective outer binary separation.

GR introduces pericenter precession of the inner binary, which can be described by the first-order
post-Newtonian (PN) theory
%%%%%%%%%%%%%%%%%%%%%%%%%%%%%%%%%%%%%%%%%%%%%%%%%%%%%%%%%%%%%%%%%%%%%%
\be\label{eq:e GR}
\frac{d \mathbf{e}_\IN}{dt}\bigg|_\gr=\dot \omega_\gr\hat{\textbf{L}}_\IN\times\mathbf{e}_\IN, ~~
\dot \omega_\gr=\frac{3Gn_\IN m_{12}}{c^2a_\IN(1-e_\IN^2)}.
\ee
%%%%%%%%%%%%%%%%%%%%%%%%%%%%%%%%%%%%%%%%%%%%%%%%%%%%%%%%%%%%%%%%%%%%%%
This precession competes with $\Omega_\lk$, and tends to suppress LK oscillations or limit
the maximum eccentricity $e_\m$ \citep[e.g.,][]{Fabrycky 2007,Liu et al 2015}.
The general secular and quasi-secular equations of motion \citep[see vector form in][]{Liu et al 2015,Liu-ApJ,Petrovich 2015},
combined with the gravitational wave (GW) radiation, completely determine the evolution of triple system.
Such LK-induced mergers have been
extensively studied \citep[e.g.,][]{Miller 2002,Blaes 2002,Wen 2003,Antonini 2012,Silsbee (2017),Liu-ApJL,Liu-ApJ,Liu spin}.

The spin vector ( $\textbf{S}_1\equiv\mathrm{S}_1 \hat{\textbf{S}}_1$)
of the BH is also coupled to the orbital angular momentum vector
$\textbf{L}_\IN$ through de-Sitter precession (1.5 PN effect) \citep[e.g.,][]{Barker 1975}:
%%%%%%%%%%%%%%%%%%%%%%%%%%%%%%%%%%%%%%%%%%%%%%%%%%%%%%%%%%%%%%%%%%%%%%
\be\label{eq:spin}
\frac{d \hat{\textbf{S}}_1}{dt}\bigg|_\mathrm{S_1L_\IN}=\Omega_\mathrm{S_1L_\IN}\hat{\mathbf{L}}_\IN \times \hat{\textbf{S}}_1, ~
\Omega_\mathrm{S_1L_\IN}=\frac{3 G n_\IN (m_{2}+\mu_\IN/3)}{2 c^2 a_\IN (1-e_\IN^2)},
\ee
%%%%%%%%%%%%%%%%%%%%%%%%%%%%%%%%%%%%%%%%%%%%%%%%%%%%%%%%%%%%%%%%%%%%%%
where $\mu_\IN\equiv m_1m_2/m_{12}$ is the reduced mass for the inner binary.
Similar equation applies to the spinning body 2.
To determine the final spin-orbit misalignments of the BHBs, it
is essential to include this spin-orbit coupling effect in the scenario of
LK-induced merger. Our recent works \citep[e.g.,][]{Liu-ApJL,Liu-ApJ,Liu spin}, focusing on the
BHB mergers induced by stellar-mass tertiary ($m_3$ comparable to
$m_1$, $m_2$), have shown that LK-induced mergers can give rise to
unique signatures for the final spin-orbit misalignment angle $\theta_{\rm sl}^{\rm f}$
\citep[see also][]{Antonini spin,Rodriguez Spin}.
In particular, for initially close BHBs (with $a_0\lesssim 0.2\au$),
which can merge without the aid of the tertiary companion,
modest ($\lesssim 40^\circ$)
$\theta_{\rm sl}^{\rm f}$ can be produced in the majority of triples \citep[e.g.,][]{Liu-ApJL}.
For wide binaries (with $a_0\gtrsim 10\au$),
the distribution of $\theta_{\rm sl}^\f$ is peaked around $90^\circ$ if the BHs have comparable masses (negligible octupole effect),
while a more isotropic distribution of final spin axis is produced
as the octupole effect increases \citep[e.g.,][]{Liu-ApJ,Liu spin}.

The {\it Standard LK-Induced Merger} mechanism, as outlined above (and studied in
all previous works), includes the key GR effects associated with the inner binaries,
but neglects the GR effects associated with the tertiary companion. This is adequate
when the tertiary mass $m_3$ is not much larger than the masses of the inner BHB.
However, for BHB-SMBH triples, with $m_3\gg m_1,m_2$, several
GR effects involving the SMBH can qualitatively change the efficiency and outcomes of LK-induced
mergers.

\section{New GR Effects Involving SMBH Tertiary}

\begin{figure}
\begin{centering}
\includegraphics[width=8.5cm]{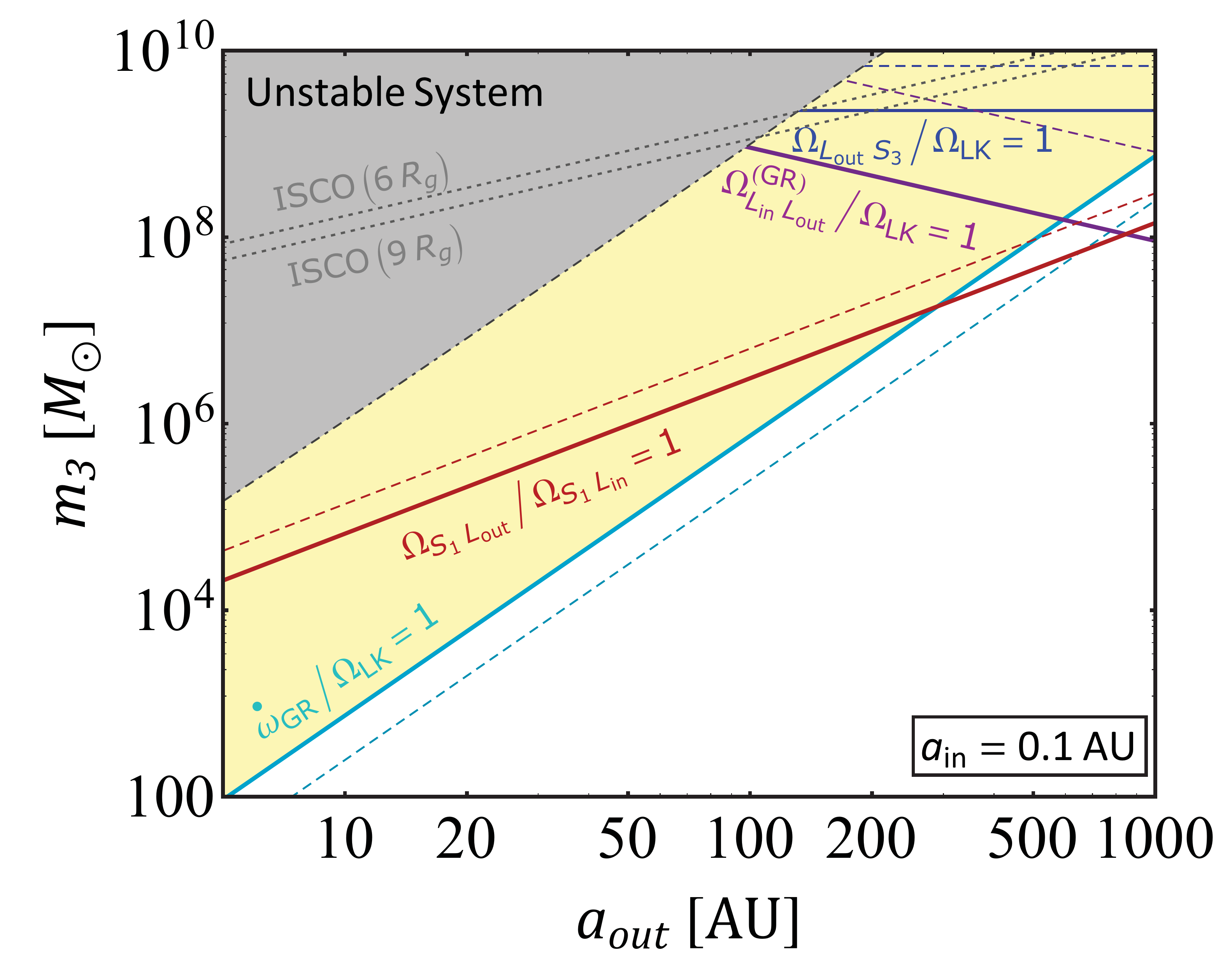}\\
\includegraphics[width=8.5cm]{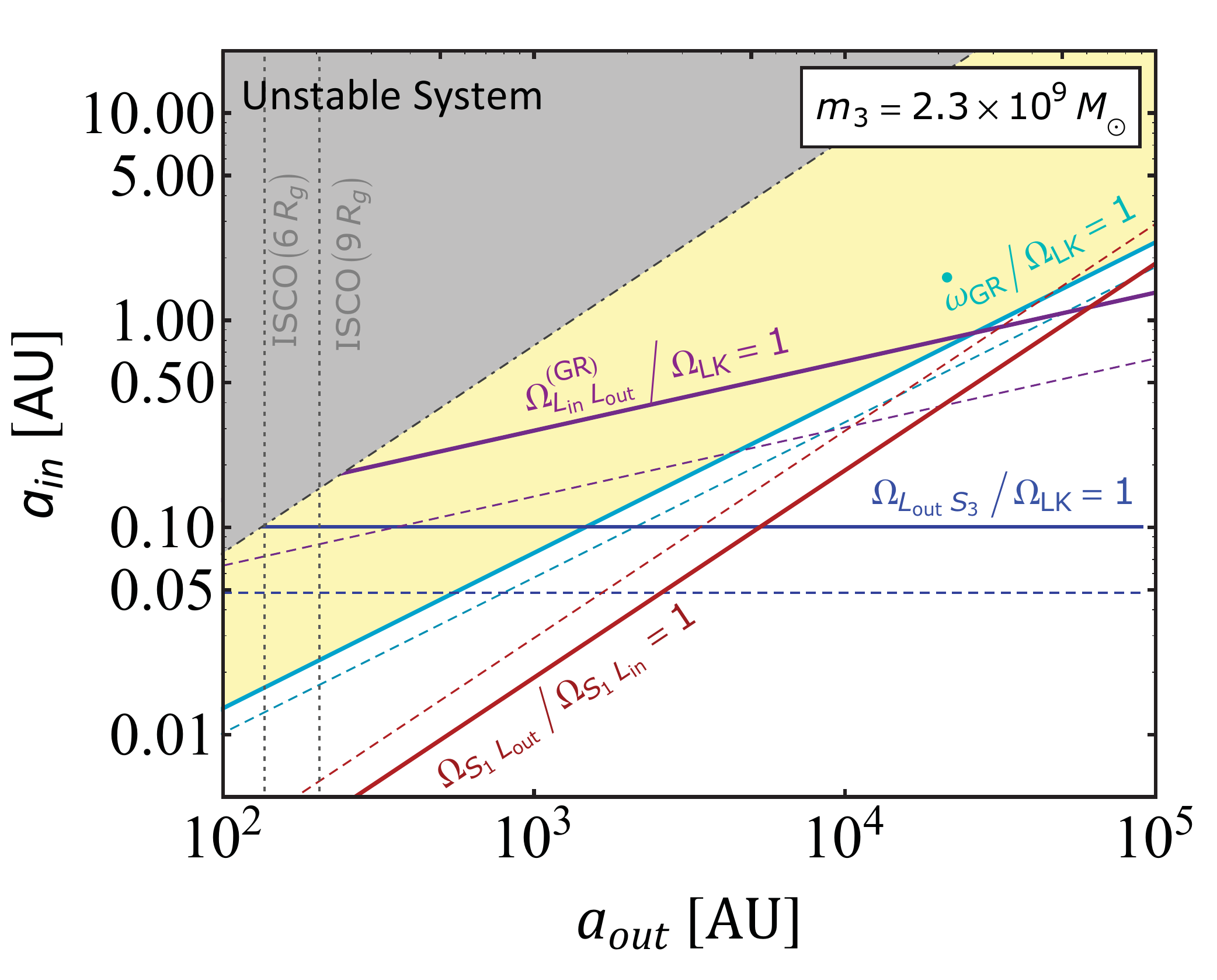}
\caption{Parameter space in the $m_3-a_\OUT$ plane and $a_\IN-a_\OUT$ plane indicating the relative importance of various GR effects.
The yellow region corresponds to the space where LK oscillations in the BHB are not suppressed by GR-induced apsidal precession
($\dot \omega_\gr/\Omega_\lk<1$) and the triple system is dynamically stable (the dot-dashed line is the instability limit
according to \cite{Kiseleva 1996}).
All the solid lines are evaluated
when the ratio of relevant frequencies is equal to unity (as labeled) and the dashed lines indicate the ratio is equal to 3.
The dotted lines indicate the innermost stable circular orbits (ISCO) for the outer binary, where $R_g=(G m_3)/c^2$
(the ISCO ranges from $R_g$ to $9R_g$ depending on the spin magnitude and orientation relative to the orbit).
The other parameters are $m_1=30M_\odot$, $m_2=20M_\odot$, $e_\IN=e_\OUT=0$ and $\chi_3=1$.
}
\label{fig:parameter space}
\end{centering}
\end{figure}

We start by examining how various GR effects associated with the SMBH tertiary
affect the LK oscillations and spin evolution of the inner BHB (see Figure \ref{fig:parameter space}).

(i)\textit{ Effect I: Lense-Thirring Precession of $\textbf{L}_\OUT$ around $\textbf{S}_3$}.
For a SMBH, the spin angular momentum $\mathrm S_3=\chi_3G m_3^2/c$
(where $\chi_3\leqslant1$ is the Kerr parameter) can be easily larger than $L_\OUT=\mu_\OUT\sqrt{Gm_\tot a_\OUT (1-e_\OUT^2)}$
[where $\mu_\OUT\equiv(m_{12}m_3)/m_\tot$ and $m_\tot=m_{12}+m_3$].
Thus $\textbf{L}_\OUT$ experiences Lense-Thirring precession around $\textbf{S}_3$ if the two vectors
are misaligned (1.5 PN effect)\citep[e.g.,][]{Barker 1975,Fang Yun}:
%%%%%%%%%%%%%%%%%%%%%%%%%%%%%%%%%%%%%%%%%%%%%%%%%%%%%%%%%%%%%%%%%%%%%%
\ba
\frac{d\textbf{L}_\OUT}{dt}\bigg|_\mathrm{L_\OUT S_3}=&&\Omega_{\mathrm{L_\OUT S_3}}\hat{\textbf{S}}_3\times\textbf{L}_\OUT\label{eq:LOUT S3}, \\
\frac{d\textbf{e}_\OUT}{dt}\bigg|_\mathrm{L_\OUT S_3}=&&\Omega_{\mathrm{L_\OUT S_3}}\hat{\textbf{S}}_3\times\textbf{e}_\OUT\nonumber\\
&&-3\Omega_{\mathrm{L_\OUT S_3}}(\hat {\textbf{L}}_\OUT\cdot \hat {\textbf{S}}_3)\hat {\textbf{L}}_\OUT\times\textbf{e}_\OUT\label{eq:EOUT S3},
\ea
%%%%%%%%%%%%%%%%%%%%%%%%%%%%%%%%%%%%%%%%%%%%%%%%%%%%%%%%%%%%%%%%%%%%%%
where the orbit-averaged precession rate is
%%%%%%%%%%%%%%%%%%%%%%%%%%%%%%%%%%%%%%%%%%%%%%%%%%%%%%%%%%%%%%%%%%%%%%
\be\label{eq:LOUT S3 rate}
\Omega_\mathrm{L_\OUT S_3}=\frac{GS_3(4+3m_{12}/m_3)}{2c^2a_\OUT^3(1-e_\OUT^2)^{3/2}}.
\ee
%%%%%%%%%%%%%%%%%%%%%%%%%%%%%%%%%%%%%%%%%%%%%%%%%%%%%%%%%%%%%%%%%%%%%%
The back-reaction of Equation (\ref{eq:LOUT S3})
implies that $\textbf{S}_3$ precesses around $\textbf{L}_\OUT$ at the
rate $\Omega_\mathrm{L_\OUT S_3}\mathrm{L_\OUT}/\mathrm{S_3}$.

As shown in \citet{Hamers Dong Quad} in a different context, the variation of
$\hat {\textbf{L}}_\OUT$ can significantly affect LK eccentricity
excitation when $\Omega_\mathrm{L_\OUT S_3}$ becomes comparable to $\Omega_{\rm LK}$.
As shown in Figure \ref{fig:parameter space}, $\Omega_\mathrm{L_\OUT S_3}/\Omega_\lk\sim 1$ can be satisfied for sufficiently
large $m_3$ ($\gtrsim 10^9M_\odot$). More precisely, LK oscillations can be
affected or triggered due to an inclination resonance, which occurs
when $\Omega_\mathrm{L_\OUT S_3}$ matches $\Omega_\mathrm{L_\IN L_\OUT}$, the precession
rate of $\hat {\textbf{L}}_\IN$ around $\hat {\textbf{L}}_\OUT$ (see below).

\begin{figure}
\begin{centering}
\includegraphics[width=9cm]{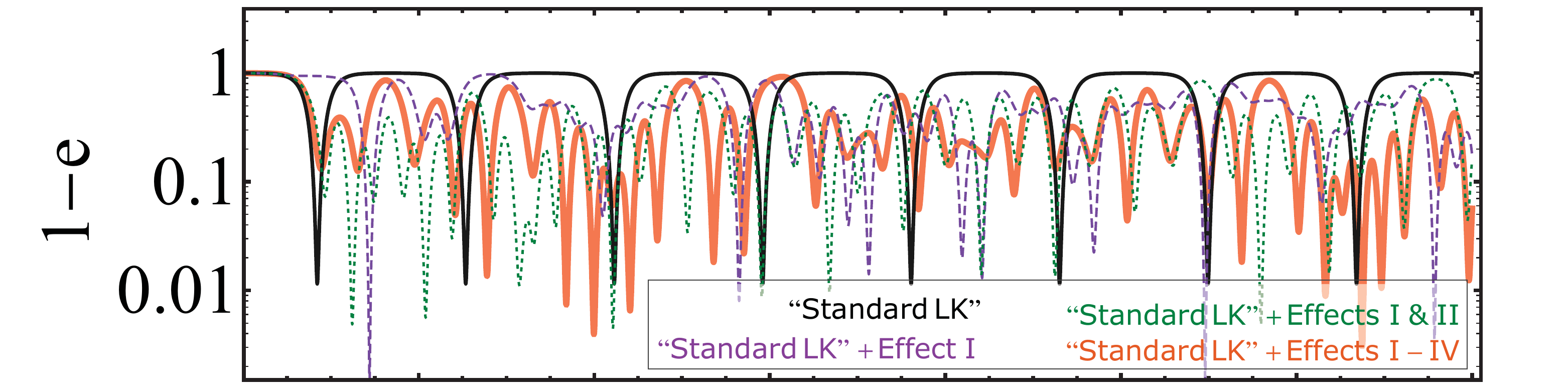}\\
\includegraphics[width=9cm]{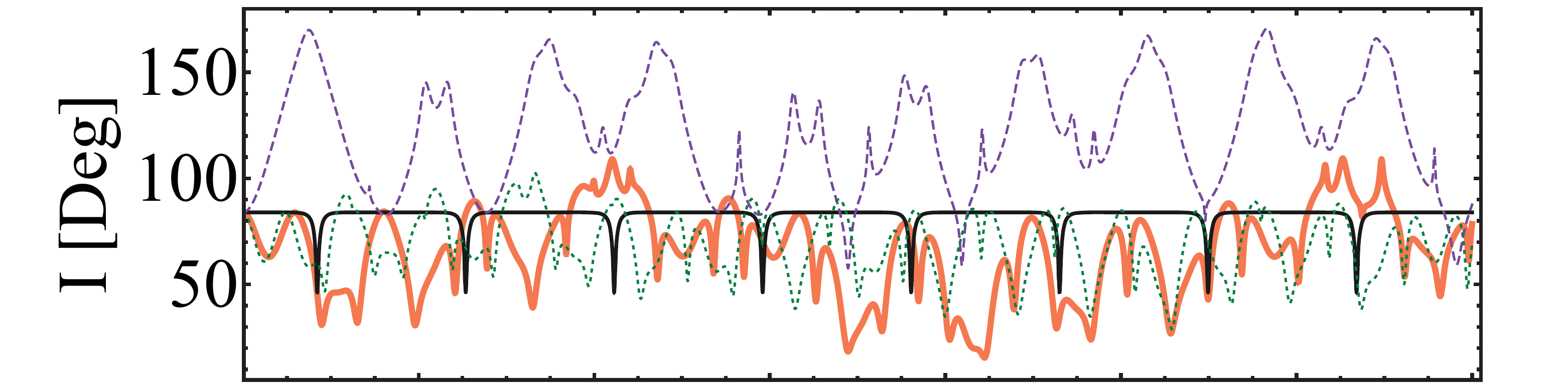}\\
\includegraphics[width=9cm]{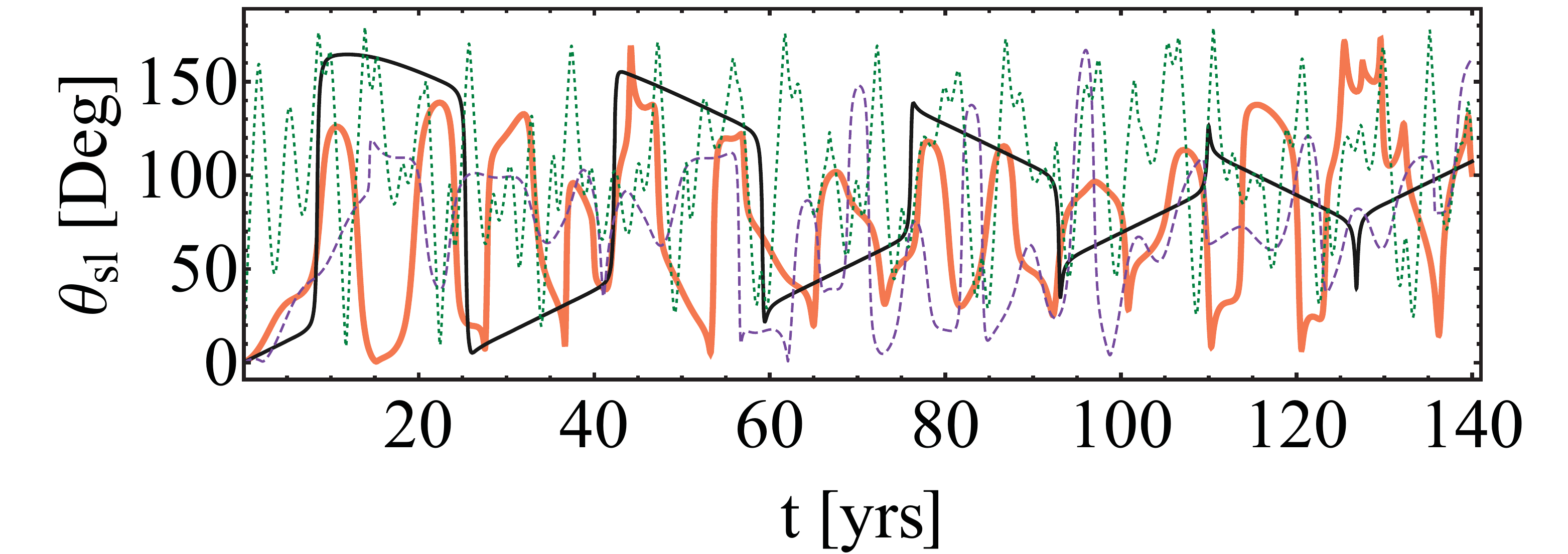}
\caption{Sample orbital and spin evolution of a BHB with a SMBH tertiary.
The three panels show the eccentricity, inclination of the inner BH binary (the angle between ${\hat{\bf L}}_\IN$ and ${\hat{\bf L}}_\OUT$), and
the spin-orbit misalignment (the angle between $\hat{\bf S}_1$ and $\hat{\bf L}_\IN$).
The parameters are $m_1=30M_\odot$, $m_2=20M_\odot$, $a_\IN=0.1\au$, $m_3=2.3\times10^9M_\odot$, $a_\OUT=500\au$, $e_\OUT=0$,
and the initial $e_{\IN,0}=0.001$, $I_0=84^\circ$ and $\theta_\SL^0=0^\circ$.
The color-coded trajectories represent the evolution with various effects included (as labeled). Gravitational radiation
is not included in these examples.
}
\label{fig:evolution}
\end{centering}
\end{figure}

Figure \ref{fig:evolution} depicts an example of how various
relativistic effects associated with the SMBH modify LK oscillations.  The results are
obtained by integrating the double-averaged (DA) secular equations of motion
\citep[averaging over both the inner and outer orbits; e.g.,][]{Liu et al 2015,Liu-ApJ}.
We see that the BHB eccentricity exhibits regular oscillations in
the ``standard LK'' case (black lines), but the inclusion of Effect I
(Equations \ref{eq:LOUT S3}-\ref{eq:EOUT S3}) (purple lines) makes the eccentricity evolve
chaotically and extend to higher values.

(ii)\textit{ Effect II: de-Sitter-like Precession of $\textbf{L}_\IN$ around $\textbf{L}_\OUT$}.
The standard LK mechanism already includes the Newtonian precession of $\textbf{L}_\IN$ around $\textbf{L}_\OUT$
(driven by the tidal potential of $m_3$ on the inner orbit), at the rate given by (to quadrupole order)
\footnote{
The general equation for finite $e_\IN$ can be found in \cite{Liu et al 2015}.
Note that for BHB-SMBH systems ($m_3\gg m_{12}$), dynamical stability requires $a_\OUT\gg a_\IN$. Thus, the
octupole LK is negligible since $\varepsilon_\oct\equiv[(m_1-m_2)/(m_1+m_2)](a_\IN/a_\OUT)[e_\OUT/(1-e_\OUT^2)]\ll1$.
}
%%%%%%%%%%%%%%%%%%%%%%%%%%%%%%%%%%%%%%%%%%%%%%%%%%%%%%%%%%%%%%%%%%%%%%
\be\label{eq:LinLout rate}
\Omega_\mathrm{L_\IN L_\OUT}^{(\mathrm{N})}=-\frac{3}{4}\Omega_\lk\big(\hat{\textbf{L}}_\OUT\cdot\hat{\textbf{L}}_\IN\big)~~~~~(\mathrm{for }~e_\IN=0).
\ee
%%%%%%%%%%%%%%%%%%%%%%%%%%%%%%%%%%%%%%%%%%%%%%%%%%%%%%%%%%%%%%%%%%%%%%
In GR, $\textbf{L}_\IN$ experiences an additional de-Sitter like (geodesic) precession
in the gravitational field of $m_3$, such that the net precession of
$\textbf{L}_\IN$ around $\textbf{L}_\OUT$ is governed by
%%%%%%%%%%%%%%%%%%%%%%%%%%%%%%%%%%%%%%%%%%%%%%%%%%%%%%%%%%%%%%%%%%%%%%
\be\label{eq:LinLout L}
\frac{d \textbf{L}_\IN}{dt}\bigg|_\mathrm{L_\IN L_\OUT}=\Omega_\mathrm{L_\IN L_\OUT}\hat{\textbf{L}}_\OUT\times\textbf{L}_\IN,
\ee
%%%%%%%%%%%%%%%%%%%%%%%%%%%%%%%%%%%%%%%%%%%%%%%%%%%%%%%%%%%%%%%%%%%%%%
with $\Omega_\mathrm{L_\IN L_\OUT}\equiv\Omega_\mathrm{L_\IN L_\OUT}^{(\mathrm{N})}+\Omega_\mathrm{L_\IN L_\OUT}^{(\gr)}$, and
%%%%%%%%%%%%%%%%%%%%%%%%%%%%%%%%%%%%%%%%%%%%%%%%%%%%%%%%%%%%%%%%%%%%%%
\be\label{eq:GR LinLout rate}
\Omega_\mathrm{L_\IN L_\OUT}^{(\gr)}=\frac{3}{2}\frac{G (m_3+\mu_\OUT/3)n_\OUT}{c^2a_\OUT(1-e_\OUT^2)},
\ee
%%%%%%%%%%%%%%%%%%%%%%%%%%%%%%%%%%%%%%%%%%%%%%%%%%%%%%%%%%%%%%%%%%%%%%
where $n_\OUT=(Gm_\tot/a_\OUT^3)^{1/2}$.
To keep $\textbf{L}_\IN\cdot\textbf{e}_\IN=0$, we also need to add
$d\textbf{e}_\IN/dt=\Omega_\mathrm{L_\IN L_\OUT}^{(\gr)}\hat{\textbf{L}}_\OUT\times\textbf{e}_\IN$ to the
eccentricity evolution equation.
We can safely neglect the feedback from $\hat{\mathbf{L}}_\IN$, $\mathbf{e}_\IN$ on $\hat{\mathbf{L}}_\OUT$ and $\mathbf{e}_\OUT$.
Equation (\ref{eq:GR LinLout rate}) has the same form as Equation (\ref{eq:spin}), but
can also be reproduced through the ``cross terms"
in the PN equations of motion of hierarchical triple systems
\citep[Private communication with Clifford Will; see also][]{C. M. Will PRD,C. M. Will PRL}.

Note that for the standard LK mechanism (and with negligible octupole effect, as valid for
the $m_3\gg m_{12}$ case considered in this paper), the nodal precession of
$\textbf{L}_\IN$ around $\textbf{L}_\OUT$ is decoupled from the LK exccentricty/inclination oscillations.
Therefore adding $\Omega_\mathrm{L_\IN L_\OUT}^{(\gr)}$ (Effect II) to
$\Omega_\mathrm{L_\IN L_\OUT}$ {\it by itself} does not alter the $e_\IN$-excitation (although it can
affect the spin evolution). However, when combined with Effect I, it can significantly affect LK
oscillation (see Figure~\ref{fig:evolution}, dotted green line).
We quantify this behavior by defining the dimensionless ratio
%%%%%%%%%%%%%%%%%%%%%%%%%%%%%%%%%%%%%%%%%%%%%%%%%%%%%%%%%%%%%%%%%%%%%%
\be\label{eq:gamma}
\gamma\equiv\frac{\Omega_\mathrm{L_\IN L_\OUT}}{\Omega_\mathrm{L_\OUT S_3}}=
\frac{\Omega_\mathrm{L_\IN L_\OUT}^{(\mathrm{N})}+\Omega_\mathrm{L_\IN L_\OUT}^{(\gr)}}{\Omega_\mathrm{L_\OUT S_3}}.
\ee
%%%%%%%%%%%%%%%%%%%%%%%%%%%%%%%%%%%%%%%%%%%%%%%%%%%%%%%%%%%%%%%%%%%%%%
Since $\Omega_\mathrm{L_\IN L_\OUT}^{(\mathrm{N})}$ depends on $I$ [where
$\hat{\textbf{L}}_\OUT\cdot\hat{\textbf{L}}_\IN=\cos I$],  $\gamma$ ranges from
$\gamma_\mi=\gamma~(I=0^\circ)$ to $\gamma_\m=\gamma~(I=180^\circ)$.

\begin{figure*}
\begin{centering}
\includegraphics[width=8.5cm]{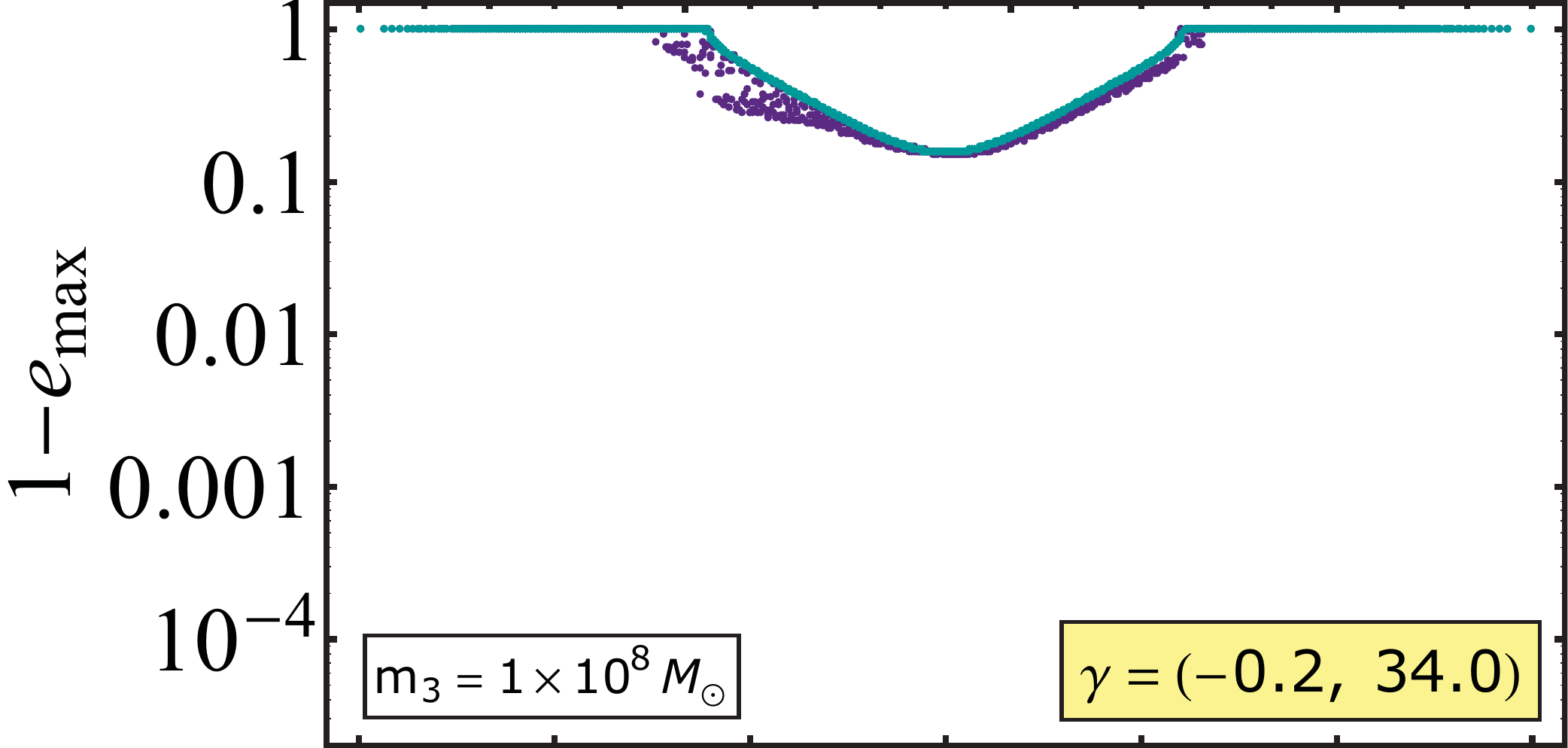}
\includegraphics[width=6.75cm]{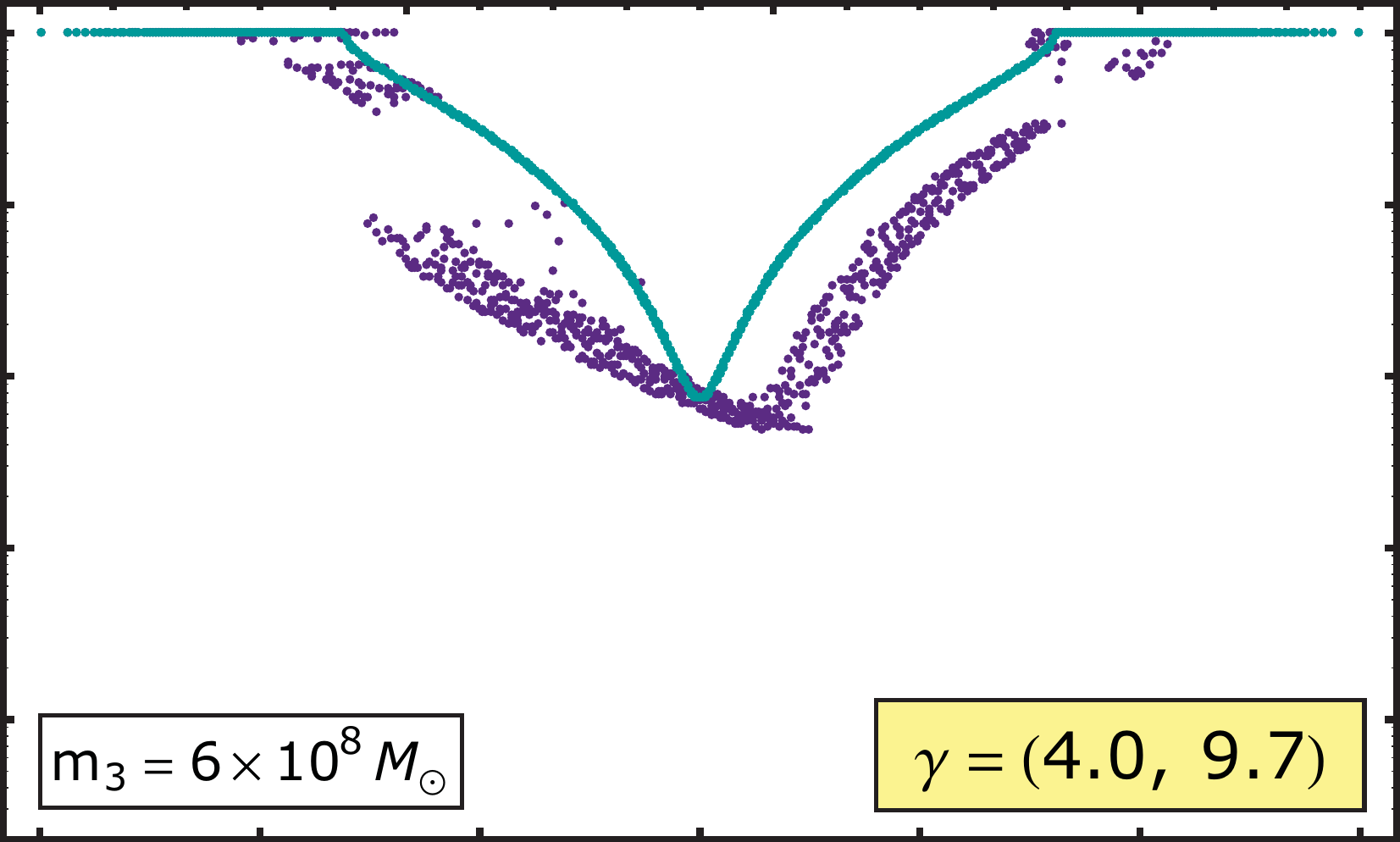}\\
\includegraphics[width=8.5cm]{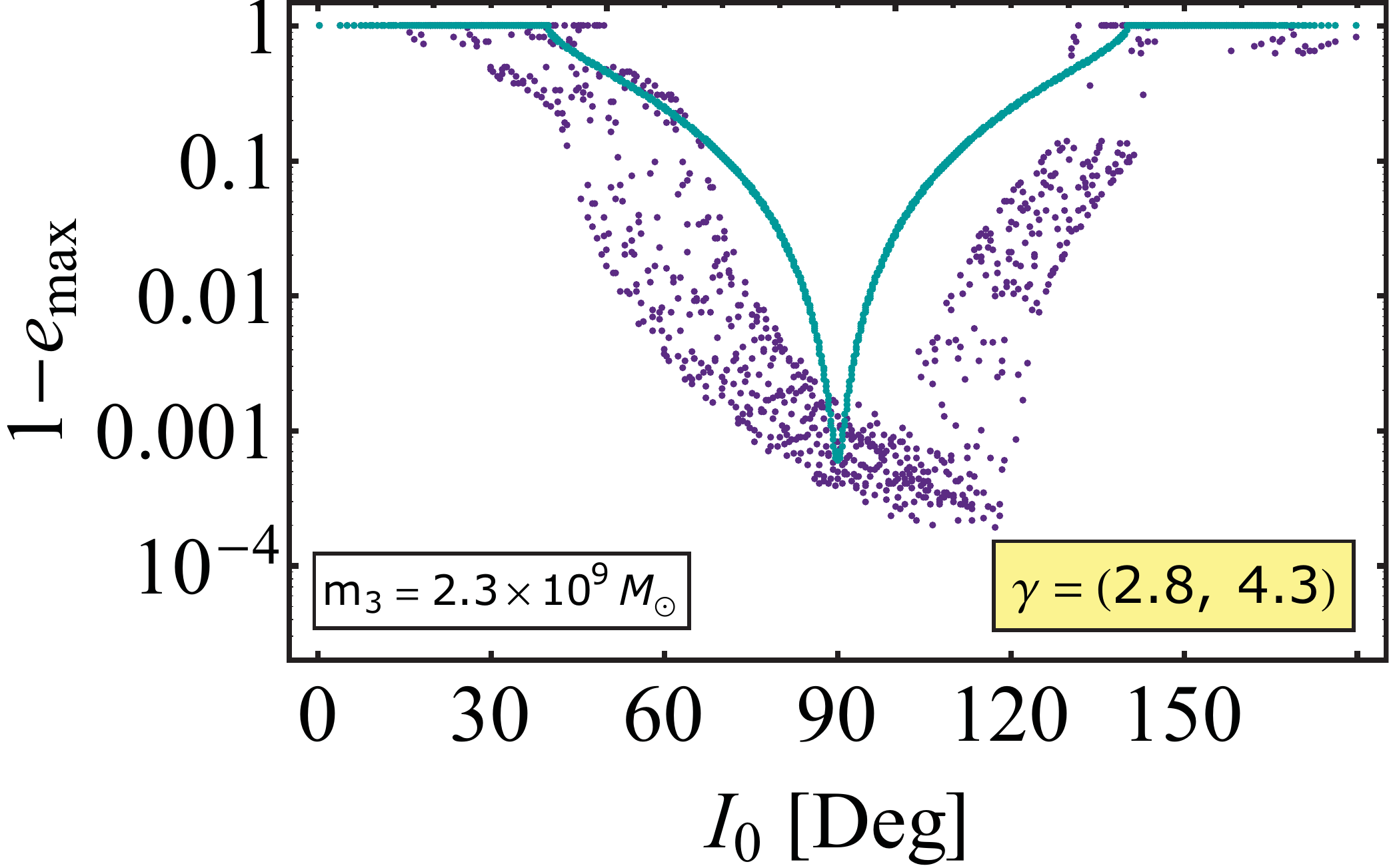}
\includegraphics[width=6.75cm]{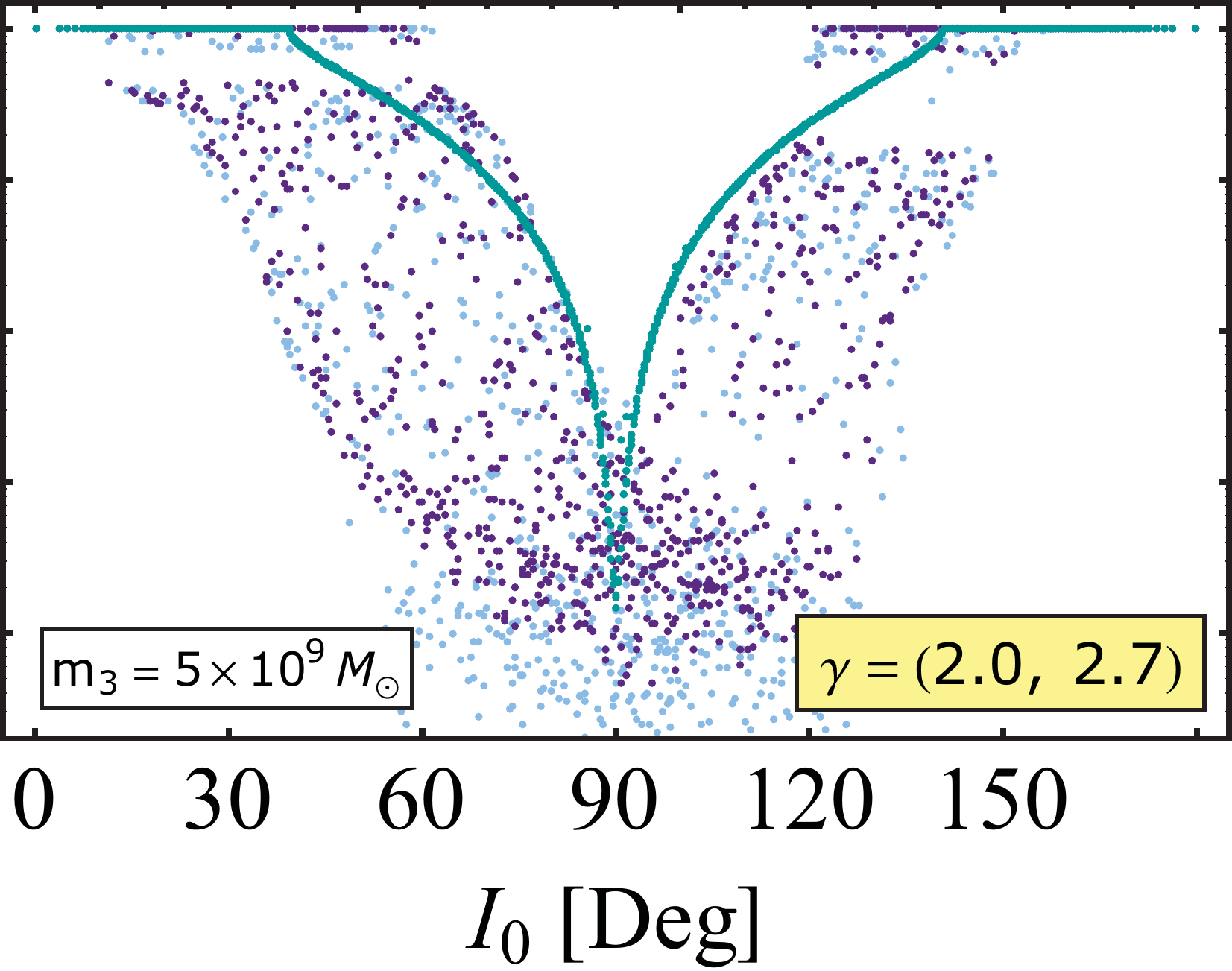}
\caption{Maximum eccentricity of the inner BHB vs. the initial inclination $I_0$
for different SMBH masses (as labeled).
The inner binary has $m_1=30M_\odot$, $m_2=20M_\odot$, $a_\IN=0.1$AU, and the SMBH has
$a_\OUT=500$AU (the initial eccentricities $e_\IN=e_\OUT=0.001$).
The misalignment angle between $\hat {\textbf{S}}_3$ and $\hat {\textbf{L}}_\OUT$ is set to $30^\circ$, but
with a random azimuthal phase angle.
The values of $e_\m$ are calculated by DA secular equations,
where the cyan dots are results \citep[which can be obtained analytically; e.g.,][]{Liu et al 2015}
from ``standard LK" and the purple dots include Effect I, II and IV.
In the bottom right panel, we also show the $e_\m$ obtained by integrating SA secular equations (light blue dots).
The range of $\gamma$ (as labeled) is given by Equation (\ref{eq:gamma}) evaluated at $I=0$ and $180^\circ$.
Effect III only influences the BH spin evolution, and thus does not play a role in the plot.
We have also done calculations that do not include Effect IV, and found that the result is similar.
}
\label{fig:LK window e excitation}
\end{centering}
\end{figure*}

As explained in \cite{Hamers Dong Quad}, when $\gamma\sim 1$,
an inclination resonance generates larger $I$ even from a small initial $I_0$, leading
to a wider range of initial inclinations for extreme eccentricity excitation.
Figure \ref{fig:LK window e excitation} explores these new GR effects by showing the
$e_\IN$-excitation window as a function of $I_0$ for BHB-SMBH systems with given $m_1,~m_2,~a_\IN,~a_{\rm out}$
but different values of $m_3$ (thus different $\gamma$'s).
The misalignment angle between $\hat {\textbf{S}}_3$ and $\hat {\textbf{L}}_\OUT$ is set to $30^\circ$, but
with a random azimuthal phase angle (i.e., the initial
$\hat{\textbf{L}}_\IN$, $\hat{\textbf{L}}_\OUT$ and $\hat{\textbf{S}}_3$ are not in the same plane
\footnote{
Note that in examples shown in \cite{Hamers Quad, Liu-Quad},
the phase angle is set to be fixed, where $\hat{\textbf{L}}_1$, $\hat{\textbf{L}}_2$
and $\hat{\textbf{L}}_\OUT$ initially lie in the same plane.
}).
By evolving the triple system using the DA secular equations,
we record $e_\m$ achieved over an integration timespan of 500 $t_\lk$
for each system with and without Effects I, II and IV.
In each panel, the cyan dots are the ``standard LK'' results; these can be calculated analytically
\citep[e.g.,][]{Liu et al 2015}. Note that since the octupole-order effects are negligible,
systems with finite $e_\OUT$ should exhibit a similar behavior as the cyan dots.
We see that including Effects I-II (purple dots) can dramatically widen the eccentricity excitation window.
As $\gamma$ approaches unity with increasing $m_3$,
overlapping inclination and LK resonances give rise to the widespread chaos \citep[e.g.,][]{Hamers Dong Quad},
causing systems with modest $I_0$ to attain extreme eccentricity growth.

When $e_\m$ becomes sufficiently close to unity,
the timescale the inner BHB spends in high-$e_\IN$ phase
\citep[$t_\lk\sqrt{1-e_\m^2}$; e.g.,][]{Anderson et al HJ} becomes less than the period of
the outer binary, the DA approximation breaks down, and the system enters semi-secular
regime \citep[e.g.,][]{Luo Liantong 2016}. If it is shorter than the inner orbital period,
the evolution of triples can only be resolved correctly by N-body integration.
In Figure~\ref{fig:LK window e excitation}, the systems in the bottom-right panel belong to
the semi-secular regime. To better address the orbital evolution, we also integrate
the single-averaged (SA) secular equations
\citep[only averaging over the inner orbital period; e.g.,][]{Liu-ApJ}.
The result (light blue dots) shows that
the eccentricity in SA integrations can undergo excursions to even more extreme values.

(iii)\textit{ Effect III: de-Sitter Precession of $\textbf{S}_1$ around $\textbf{L}_\OUT$}.
The ``standard LK" already includes de-Sitter precession of $\textbf{S}_1$ around $\textbf{L}_\IN$.
With a SMBH tertiary, $\textbf{S}_1$ also experiences a precessional torque from $m_3$:
%%%%%%%%%%%%%%%%%%%%%%%%%%%%%%%%%%%%%%%%%%%%%%%%%%%%%%%%%%%%%%%%%%%%%%
\be\label{eq:spin Lout}
\frac{d \hat{\textbf{S}}_1}{dt}\bigg|_\mathrm{S_1L_\OUT}=\Omega_\mathrm{S_1L_\OUT}\hat{\mathbf{L}}_\OUT \times \hat{\textbf{S}}_1,
\ee
%%%%%%%%%%%%%%%%%%%%%%%%%%%%%%%%%%%%%%%%%%%%%%%%%%%%%%%%%%%%%%%%%%%%%%
with
%%%%%%%%%%%%%%%%%%%%%%%%%%%%%%%%%%%%%%%%%%%%%%%%%%%%%%%%%%%%%%%%%%%%%%
\be\label{eq:spin Lout rate}
\Omega_\mathrm{S_1L_\OUT}=\frac{3}{2}\frac{G (m_3+\mu_\OUT/3)n_\OUT}{c^2a_\OUT(1-e_\OUT^2)}.
\ee
%%%%%%%%%%%%%%%%%%%%%%%%%%%%%%%%%%%%%%%%%%%%%%%%%%%%%%%%%%%%%%%%%%%%%%
Note that $\Omega_\mathrm{S_1L_\OUT}=\Omega_\mathrm{L_\IN L_\OUT}^{(\gr)}$ (Equation~\ref{eq:GR LinLout rate}).
The back-reaction torques on $\hat{\mathbf{L}}_\OUT$ and
$\hat{\mathbf{e}}_\OUT$ can be safely neglected since
$\mathrm{L}_\OUT\gg\mathrm{S}_1$. Although Equation~(\ref{eq:spin Lout})
does not affect the orbital evolution of the inner binary, it does
affect the evolution of $\textbf{S}_1$ and the spin-orbit misalignment angle $\theta_\SL$.

The bottom panel of Figure~\ref{fig:evolution} shows several examples of the evolution
of $\theta_\SL$ during LK oscillations, with and without various
GR effects. The evolution of ${\bf S}_1$ is governed by two
``adiabaticity parameters":
%%%%%%%%%%%%%%%%%%%%%%%%%%%%%%%%%%%%%%%%%%%%%%%%%%%%%%%%%%%%%%%%%%%%%%
\be\label{eq:adiabaticity parameter}
\mathcal{A}\equiv \left|\frac{\Omega_\mathrm{S_1L_\IN}}{\Omega_\mathrm{L_\IN L_\OUT}}\right|,
%\simeq\frac{\Omega_\mathrm{S_1L_\IN}}{\Omega_\lk\Big[1+\Omega_\mathrm{L_\IN L_\OUT}^{(\gr)}/\Omega_\lk\Big]},
~~~\mathcal{B}\equiv\frac{\Omega_\mathrm{S_1L_\IN}}{\Omega_\mathrm{S_1L_\OUT}}.
\ee
%%%%%%%%%%%%%%%%%%%%%%%%%%%%%%%%%%%%%%%%%%%%%%%%%%%%%%%%%%%%%%%%%%%%%%
We expect (i) When $\mathcal{A}, \mathcal{B}\ll 1$ (``nonadiabatic''),
the spin axis $\hat{\bf S}_1$ cannot ``keep up'' with the rapidly changing
$\hat{\textbf{L}}_\IN$, and thus effectively precesses around $\textbf{L}_\OUT$, keeping
$\theta_\mathrm{S_1L_\OUT}\simeq$ constant [Note that since
$\Omega_\mathrm{S_1L_\OUT}=\Omega_\mathrm{L_\IN L_\OUT}$ is only a few times larger than
$\Omega_\mathrm{L_\OUT S_3}$ (see Figure~\ref{fig:parameter space}), $\theta_\mathrm{S_1L_\OUT}$
is only approximately constant as $\hat{\textbf{L}}_\OUT$ precesses around $\hat{\textbf{S}}_3$];
(ii) When $\mathcal{A}, \mathcal{B}\gg 1$ (``adiabatic''),
$\hat{\textbf{S}}_1$ closely ``follows'' $\hat{\textbf{L}}_\IN$, maintaining an approximately constant $\theta_\SL$.
(iii) In the regime between (i) and (ii) (``trans-adiabatic''), the
evolution of $\hat{\textbf{S}}_1$ can be quite complicated and chaotic, because of
its dependence on $e_\IN$ during the LK cycles
\citep[see][]{Dong Science,Storch 2015,Anderson et al HJ,Anderson et al 2017,Liu-ApJL,Liu-ApJ}.

As the BHB orbit decays, the system may transitions from ``nonadiabatic" at large $a_\IN$ to ``adiabatic"
at small $a_\IN$, where the final spin-orbit misalignment angle $\theta_\SL^\f$ is ``frozen".
From Figure~\ref{fig:parameter space}, we see that, because of the contribution of
$\Omega_\mathrm{L_\IN L_\OUT}^{(\gr)}$ to $\Omega_\mathrm{L_\IN L_\OUT}$,
the conditions $\mathcal{A}, \mathcal{B}\ll1$ can be easily satisfied initially for systems with $m_3\gtrsim10^8M_\odot$.
As these systems experience LK-induced orbital decay,
they must go through the ``trans-adiabatic" regime and therefore may attain a wide range of
$\theta_\SL^\f$ (see below).

(iv)\textit{ Effects IV}.
Both $\hat {\textbf{L}}_\IN$ and
$\hat {\textbf{S}}_1$ (and $\hat {\textbf{S}}_2$) experience Lens-Thirring precession
around $\hat {\textbf{S}}_3$ at the rate
%%%%%%%%%%%%%%%%%%%%%%%%%%%%%%%%%%%%%%%%%%%%%%%%%%%%%%%%%%%%%%%%%%%%%%
\be\label{eq:LT}
\Omega_\mathrm{L_\IN S_3}=\Omega_\mathrm{S_1S_3}=\Omega_\mathrm{LT}=\frac{GS_3}{2c^2a_\OUT^3(1-e_\OUT^2)^{3/2}}.
\ee
%%%%%%%%%%%%%%%%%%%%%%%%%%%%%%%%%%%%%%%%%%%%%%%%%%%%%%%%%%%%%%%%%%%%%%
Since $\Omega_\mathrm{L_\IN S_3}/\Omega_\mathrm{L_\IN L_\OUT}^{(\gr)}
=\Omega_\mathrm{S_1S_3}/\Omega_\mathrm{S_1L_\OUT}\sim V_\OUT/c$ (where $V_\OUT$ is the orbital velocity of the outer binary),
they can be neglected when $V_\OUT/c\ll1$.

\section{Binary BH Mergers Induced by SMBH}

\begin{figure}
\begin{centering}
\includegraphics[width=8cm]{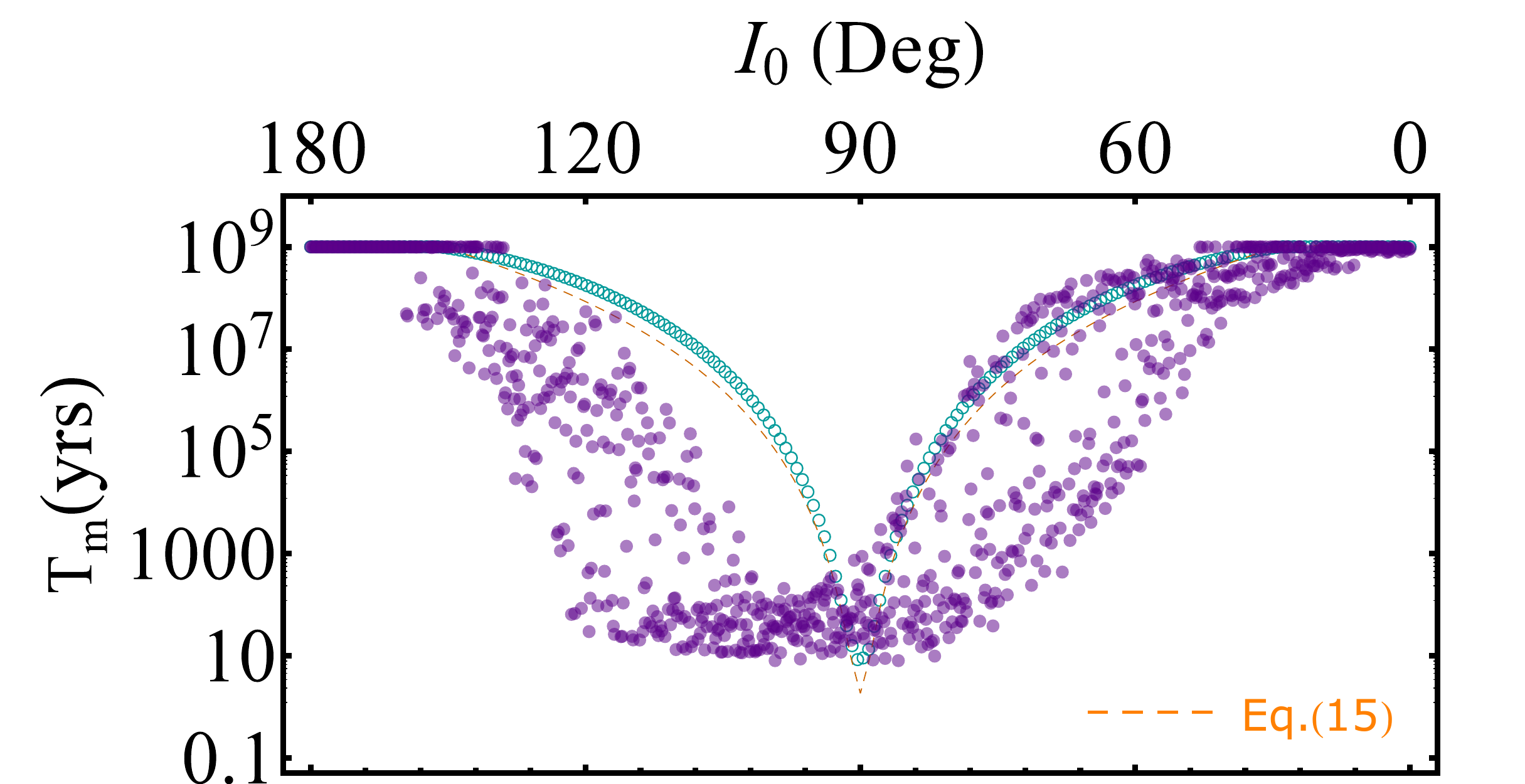}\\
\includegraphics[width=8cm]{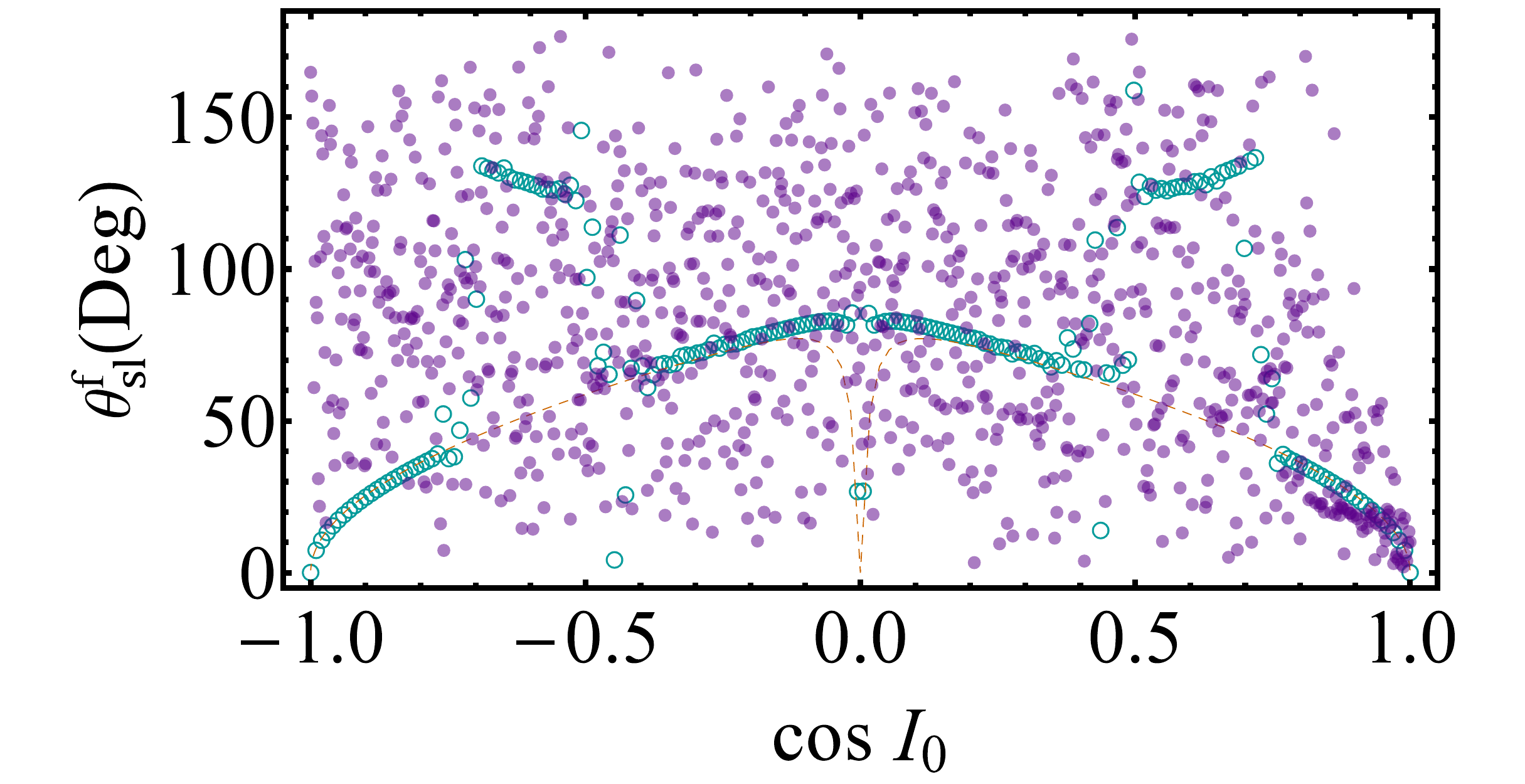}\\
\includegraphics[width=8cm]{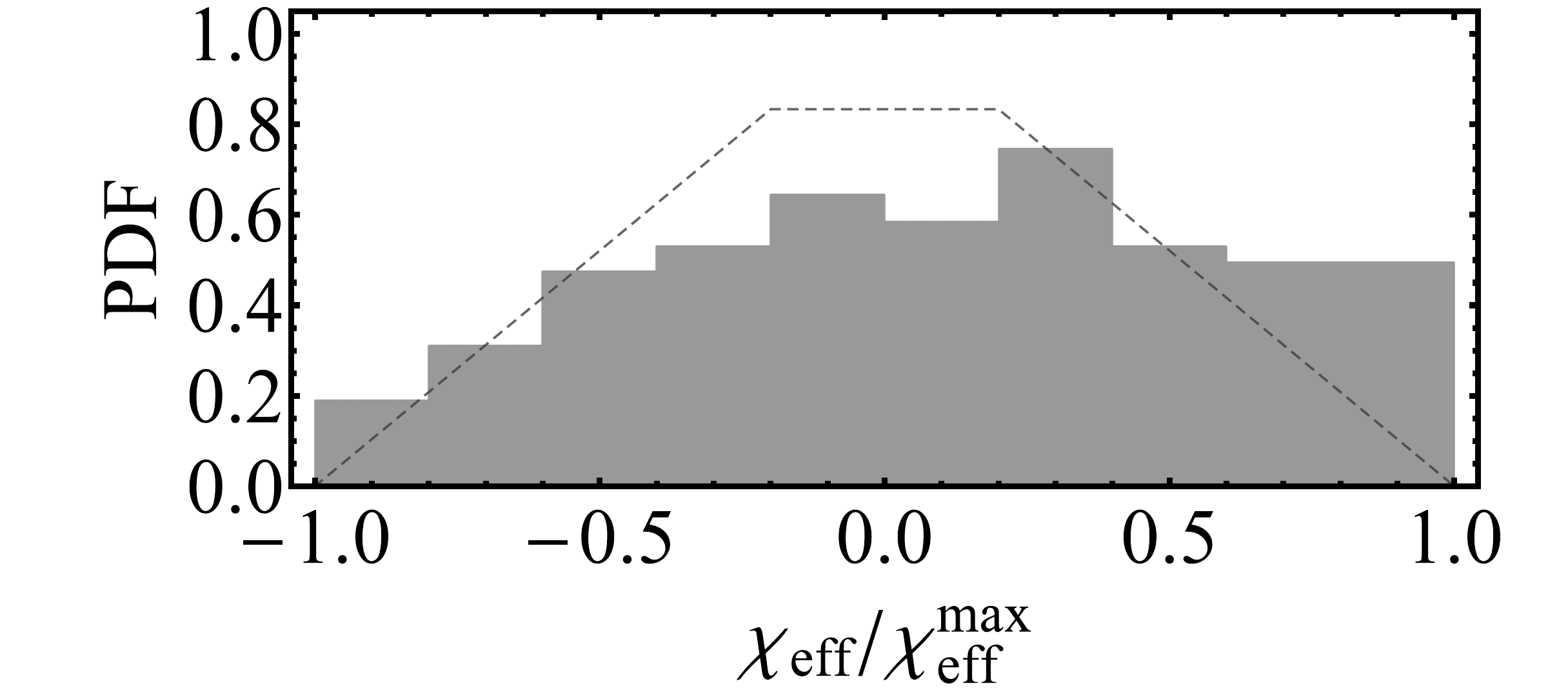}
\caption{The BHB merger time $T_\mathrm{m}$ (top panel) and final spin-orbit misalignment angle (middle panel)
as a function of the initial inclination for the BHB-SMBH triple system.
The purple dots are the results that include various new GR effects
discussed in this paper (Effects I-III), while the cyan circles do not.
The bottom panel shows the distribution of the rescaled binary spin parameter $\chi_\eff$
[with $\chi_\eff^\m=(m_1\chi_1+m_2\chi_2)/m_{12}$ and assuming $\chi_1=\chi_2$]
for the ``GR-enhanced" mergers (purple dots in the moddle panel).
The system parameters are the same as Fig.~\ref{fig:LK window e excitation}
with $m_3=2.3\times10^9M_\odot$. The dashed curve in the middle panel is given by the analytical
expression derived for circular mergers in the presence of a tertiary \cite{Liu-ApJL}, and the dashed line in the bottom panel shows the distribution
for uncorrelated isotropic spins (Eq. 81 in \cite{Liu-ApJ}).
}
\label{fig:merger window}
\end{centering}
\end{figure}

We now add gravitational radiation in our fiducial example (Figure~\ref{fig:LK window e excitation} with
$m_3=2.3\times10^9M_\odot$). Since the Effect IV is not important in this example, we
perform two sets of calculations with and without Effects I-III,
evolve the system until the BHB enters the LIGO band (i.e., when the peak GW frequency reaches 10~Hz).
The results are summarized in Figure~\ref{fig:merger window}.

In the ``standard LK'' mechansim (without Effects I-III; cyan circles in the top two
panels of Figure~\ref{fig:merger window}), for systems with negligible octupole effects,
the merger time can be well approximated by \citep[e.g.,][]{Liu-ApJ}
%%%%%%%%%%%%%%%%%%%%%%%%%%%%%%%%%%%%%%%%%%%%%%%%%%%%%%%%%%%%%%%%%%%%%%
\be\label{eq:fitting formula}
T_\mathrm{m}\simeq T_{\mathrm{m},0}(1-e_\m^2)^3,
\ee
%%%%%%%%%%%%%%%%%%%%%%%%%%%%%%%%%%%%%%%%%%%%%%%%%%%%%%%%%%%%%%%%%%%%%%
where $T_{\mathrm{m},0}\equiv(5c^5 a_{\IN,0}^4)/(256 G^3 m_{12}^2 \mu_\IN)$ is the merger time due to GW emission
for an isolated circular BHB \citep[e.g.,][]{Peters 1964} ($T_{\mathrm{m},0}\simeq 10^9$~yrs for the systems
considered in Figure~\ref{fig:merger window}), and $e_{\mathrm{max}}$ is the maximum eccentricity achieved in the LK cycle
(see Figure \ref{fig:LK window e excitation}).
When the GR effects associated with the SMBH are taken into account (purple dots), the range of inclinations for rapid mergers (shorter $T_{\rm m}$)
becomes much larger, a direct consequence of the widened LK eccentricity
excitation window (see Figure~\ref{fig:LK window e excitation}).
Note that in a dense nuclear cluster, the orbits of a BHB-SMBH triple system
can be perturbed or disrupted by close fly-bys of other objects.
If we introduce upper limits of the survival time for the triples,
the ``standard LK'' would give the merger fraction of $f_\merger\simeq 12\%,20\%,30\%$ for $T_\mathrm{m}
\lesssim 10^5,10^6,10^7$~yrs, respectively, while including Effects I-III would increase
the corresponding merger fraction to $f_\merger\simeq58\%,63\%,70\%$.

The middle panel of Figure \ref{fig:merger window} shows the distribution of $\theta_\SL^\f$ as a function of $\cos I_0$
\footnote{
In a nuclear cluster, the initial binary BHs may have nontrivial spin orientations due to
the complicated scattering processes. In order to have an intuitive understanding
of the spin dynamics, here we assume that the BH spin axis is initially aligned with the orbital axis.}.
In the ``standard LK'' (as studied in \cite{Liu-ApJL,Liu-ApJ,Liu spin}), the final spin axis shows a
regular distribution when the octupole effects are negligible (as in the BHB-SMBH case studied here);
for the systems that do
not experience eccentricity excitation,
an analytical expression for $\theta_\SL^\f$ can be obtained \citep[][]{Liu-ApJL} (see the dashed line).
However, when the GR effects associated with the SMBH are included, the final BH spin
orientation is significantly ``randomized''.
Given the wide distribution of $\theta_\SL^\f$, we find the large spread in $\chi_\eff$ in the bottom panel
of Figure~\ref{fig:merger window}, where $\chi_{\rm eff}=
(m_1 \boldsymbol{\chi}_1+m_2 \boldsymbol{\chi}_2)\cdot\hat{\mathbf{L}}_\IN/m_{12}$
[with $\boldsymbol{\chi}_{1,2}=c\mathbf{S}_{1,2}/(Gm_{1,2}^2)$]
is the effective binary spin parameter that can be directly measured from GW observations.
Note that the two spins in the merging binary BHs are strong correlated (see also \cite{Liu spin}; Fig. 10);
this is different for the scenarios involving strong scattering,
which expectedly produce uncorrelated isotropic spins.

Due to the negligible octupole effect in BHB-SMBH systems, the ``residual" eccentricities of merging BHBs
(when they enter the 10 Hz LIGO band) are all below $0.1$ in our simulations.
This is in contrast to binary mergers induced by a stellar-mass tertiary studied in \cite{Liu spin}.

\section{Summary and Discussion}

We have identified the impacts of several GR effects in BHB-SMBH triples that
have been little explored.
Effect I (Equations \ref{eq:LOUT S3}-\ref{eq:LOUT S3 rate})
allows the BHB eccentricity to reach extremely high values even with modestly inclined or nearly coplanar outer orbits.
Effect II (Equations \ref{eq:LinLout L}, \ref{eq:GR LinLout rate}) modifies the eccentricity growth (when combined with Effect I)
and BH spin evolution indirectly.
Effect III (Equations \ref{eq:spin Lout}, \ref{eq:spin Lout rate}) only affects the spin evolution.
The overall dynamics of the BHB and BH spin around a SMBH
can be characterized by the dimensionless rates (Equations \ref{eq:gamma}, \ref{eq:adiabaticity parameter}).
Effects I and II generally require very massive SMBH ($m_3 \gtrsim 10^8-10^9M_\odot$) to be effective,
while Effect III can be important for a wide range of SMBH masses (see Figure \ref{fig:parameter space}). Overall, these
GR effects can significantly widen the LK-induced merger window and increase the merger fraction. They
also produce a broad distribution of the final BH spin-orbit misalignment angles,
leading to a wide range of the effective BHB spin parameter $\chi_\eff$.

Our proof-of-concept calculations have demonstrated the importance of the GR effects in BHB-SMBH systems.
However, we have not thoroughly explored the relevant parameter space, nor considered various ``environmental" effects associated with BHBs
in nuclear cluster. We leave these to future works.

\section{Acknowledgments}

We thank Jean Teyssandier and Clifford Will for useful discussion and communication.
This work is supported in part by the NSF grant AST-1715246 and NASA
grant NNX14AP31G.

\end{document}